\documentstyle[twoside,fleqn,espcrc2,epsf,rotate]{article}

\newcommand{\AmS}{{\protect\the\textfont2
  A\kern-.1667em\lower.5ex\hbox{M}\kern-.125emS}}

\def\Preprint{\vspace*{-8.5cm} 
 \noindent FTUV/97-46 \\ 
  IFIC/97-47 \\  HD-THEP-97-50 \\ 
  \vspace{5.8cm}}
\def\refjl#1#2#3#4#5#6{\bibitem{#1} #2, {#3} {#4} (#5) #6.}
\def\refbk#1#2#3#4{\bibitem{#1} #2, {\it #3}, #4.}
\def\etal{{et al}}
\def\NP{Nucl. Phys.}

\def\PL{Phys. Lett.}
\def\PRL{Phys. Rev. Lett.}
\def\PR{Phys. Rev.}
\def\PRep{Phys. Rep.}
\def\ZP{Z. Phys.}

\def\MP{Int. J. Mod. Phys.}

\newcommand{\eqn}[1]{(\ref{#1})}
\newcommand{\be}{\begin{equation}}
\newcommand{\ee}{\end{equation}}
\newcommand{\no}{\nonumber}
\newcommand{\bel}[1]{\be\label{#1}}
\newcommand{\ba}{\begin{array}{c}}
\newcommand{\bat}{\begin{array}{cc}}
\newcommand{\ea}{\end{array}}
\newcommand{\beqn}{\begin{eqnarray}}
\newcommand{\eeqn}{\end{eqnarray}}

\newcommand{\bi}{\begin{itemize}}
\newcommand{\ei}{\end{itemize}}

\newcommand{\gsim}{~{}_{\textstyle\sim}^{\textstyle >}~}

\newcommand{\cM}{{\cal M}}
\newcommand{\cO}{{\cal O}}

\newcommand{\Mn}{{\cal M}_n}
\newcommand{\wt}{\widetilde}
\newcommand{\as}{\alpha_{s}}

\newcommand{\IM}{\mbox{\rm Im}}

\newcommand{\GG}{\big<aGG\big>}

\newcommand{\mev}{\mbox{\rm MeV}}
\newcommand{\gev}{\mbox{\rm GeV}}

\newcommand{\MSb}{{\overline{MS}}}

\title{Determination of $M_b$ and $\alpha_s$ from the
  Upsilon System}

\author{M. Jamin\address{
         Institut f\"ur Theoretische Physik, Universit\"at Heidelberg, \\
         Philosophenweg 16, D-69120 Heidelberg, Germany}
        and
        A. Pich\address{
         Departament de F\'{\i}sica Te\`orica, 
         IFIC,  CSIC --- Universitat de Val\`encia, \\ 
         Dr. Moliner 50, E--46100 Burjassot, Val\`encia, Spain\protect\thanks{
    Invited talk at the High Energy Conference on Quantum Chromodynamics
    (QCD'97), Montpellier, July 1997}}
}
       
\begin{document}

\begin{abstract}
The mass of the bottom quark 
(both the pole mass $M_b$ and the $\MSb$ mass $m_b$)
and the strong coupling constant $\alpha_s$ 
have been determined \protect\cite{JP:97}
from QCD moment sum rules for the $\Upsilon$ system.
In the pole--mass scheme large perturbative corrections
resulting from coulombic contributions have been resummed.
The results of this analysis are:
$M_b=4.60 \pm 0.02\,\gev$, $m_b(m_b)=4.13 \pm 0.06\,\gev$
and $\alpha_s(M_Z)=0.119 \pm 0.008$.

\end{abstract}

\maketitle
\Preprint

\section{INTRODUCTION}
\label{sec:introduction}

A short--distance description in terms of quarks and gluons is
well suited for inclusive quantities, where no reference to a particular
hadronic state is needed. The vacuum polarization
$$
\left(q_\mu q_\nu-g_{\mu\nu}q^2\right) \Pi(q^2)  =  i \!\!\int \!\! dx \, 
e^{iqx} \langle  T\{j_\mu(x) j_\nu(0)\} \rangle 
$$
induced
by the vector current 
$j_\mu\equiv\bar b\gamma_\mu b$
can be calculated theoretically within the Operator Product Expansion
(OPE), whereas its imaginary
part can be experimentally determined from the $e^+e^-\to b \bar b$
cross-section:
$$
R(s)\equiv {1\over Q_b^2}
\frac{\sigma(e^+e^-\to b \bar b)}{\sigma(e^+e^-\to\mu^+\mu^-)}
\ =  12\pi \,\IM\,\Pi(s+i\epsilon) \, .
$$

Using a dispersion relation the $n$th derivative of $\Pi(s)$ at $s=0$ can
be expressed in terms of the $n$th integral moment of $R(s)$:
\beqn\label{eq:M_def}
\Mn &\!\!\! \equiv &\!\!\!
\frac{12\pi^2}{n!}\left.\left(4M_b^2\,\frac{d}{ds}
\right)^n \Pi(s)\right\vert_{s=0} 
\no\\
&\!\!\! = &\!\!\! (4M_b^2)^n \int\limits_0^\infty \!
ds \, \frac{R(s)}{s^{n+1}} 
\\ &\!\!\! = &\!\!\!
2 \int\limits_0^1 \! dv \, v(1-v^2)^{n-1} R(v) \, , \no
\eeqn
where $v\equiv\sqrt{1-4M_b^2/s}$.
$M_b$ corresponds to the pole of the perturbatively renormalized propagator,
whereas the running quark mass in the $\MSb$ scheme 
renormalized at a scale $\mu$ will be denoted by
$m_b(\mu)$.

Under the assumption of quark--hadron duality, the moments $\Mn$ can either
be calculated theoretically in renormalization group improved perturbation
theory, including non-perturbative condensate contributions, or can be
obtained from experiment. In this way, hadronic quantities like masses
and decay widths get related to the QCD parameters $\alpha_s$, $m_b$ and
condensates.

For large values of $n$, the moments become dominated by the threshold
region. Therefore, they are very sensitive to the heavy quark mass.
Owing to the large size of the $(\alpha_s \sqrt{n})^k$ Coulombic
corrections, the large--$n$
moments can also be used to get a determination of $\alpha_s$
from the existing data on $\Upsilon$ resonances \cite{vol:95}.

\section{EXPERIMENTAL INPUT}

The first six Upsilon resonances have been observed. Their
masses are known rather accurately, and their electronic widths
have been measured with an accuracy which ranges from
3\% for the $\Upsilon(1S)$ to 23\% for the $\Upsilon(6S)$.
For our purposes, 
the narrow--width approximation provides a very good description of
these states, because
the full widths of the first three $\Upsilon$ resonances are roughly a factor
$10^{-5}$ smaller than the corresponding masses, and
the higher--resonance contributions to the moments are suppressed:
\begin{equation}
\label{eq:6.2}
{\Mn\over (4M_b^2)^n} =  \frac{9\pi}{\bar\alpha^2 Q_b^2}\,
\sum\limits_{k=1}^6 \frac{\Gamma_{kS}}{M_{kS}^{2n+1}} +
\int\limits_{s_0}^\infty \!ds \,\frac{R(s)}{s^{n+1}} \,,
\end{equation} 
where 
$\Gamma_{kS}\equiv\Gamma[\Upsilon(kS)\to e^+e^-]$
and \cite{pdg:96} $\bar\alpha^2=1.07\,\alpha^2$.

The $e^+e^-\to b\bar b$ cross-section above threshold
is unfortunately very badly measured  \cite{cleo:91}.
The second term in Eq.~\eqn{eq:6.2} accounts for the contributions to
$R_b$ above the sixth resonance and is approximated by the perturbative
QCD continuum. 
The continuum threshold $\sqrt{s_0}$ should lie 
around the mass of the next resonance, which has been estimated in potential
models.  
We have used
$\sqrt{s_0}=11.2\pm0.2\,\gev$; the lower value includes the mass of
the sixth resonance and should be a conservative estimate. The
contribution from open $B$ production above the $B\bar B$
threshold and below $s_0$ is very small \cite{cleo:91}
and has been included in the variation of $s_0$.

The numerical weight of the heavier resonances in~\eqn{eq:6.2}
decreases strongly for increasing values of $n$. The contribution of the
$\Upsilon(5S)$ [$\Upsilon(6S)$] is 9.5\% [4\%] at $n=0$; 1\% [0.3\%] at
$n=10$; and a tiny 0.08\% [0.02\%] at $n=20$. Therefore, taking $n\gsim 10$,
the uncertainties associated with the contributions of higher--mass states 
are very small.

\section{THEORETICAL CALCULATION}

\subsection{Perturbation Theory}

The vacuum polarization $\Pi(s)$ can be expanded in
powers of the strong coupling constant:
$$
\Pi(s)  =  \Pi^{(0)}(s)+a\,\Pi^{(1)}(s)+a^2\,\Pi^{(2)}(s)+\ldots \,,
$$
with $a\equiv\as/\pi$. Analogous expansions for $R(v)$
and $\Mn$ can be written.
For the first two terms, analytic expressions are available
\cite{ks:55,schw:73}.    

$\Pi^{(2)}(s)$ is still not fully known
analytically. 
 However, the method of Pad\'e approximants has been recently exploited to
calculate $\Pi^{(2)}$ numerically, using available results at
high energies,          
analytical results for the first seven
moments ${\cal M}_i^{(2)}$ ($i=1,\ldots,7$) and the known threshold behaviour
of $R^{(2)}(v)$ \cite{bb:95,cks:96}.      
This information is good enough to obtain an accurate numerical evaluation
of $\cM^{(2)}_n$ for values of $n$ not too large.

The contributions from diagrams with internal quark
loops to the spectral function are fully known.  
This allows to check the accuracy of the Pad\'e approximation, which
for $n=20$ is found to be better than $10^{-6}$.

The numerical stability of the results can be analyzed, either
using different Pad\'e approximants with the full set of
information, or constructing Pad\'e approximants with one order less
by removing one datum.
For $n\leq 20$, the resulting numerical uncertainty is below 0.02\%,
being completely negligible for our application.

The reliability of the Pad\'e approximation has been further corroborated
in Ref.~\cite{ChHKS:97},
where the first seven terms of the expansion of $\Pi^{(2)}(s)$ in
powers of $M_b^2/s$ have been computed. 


\subsection{Coulomb Resummation}

For large $n$ the higher--order perturbative corrections to the moments
grow with respect to the leading order. 
At $n=8$ ($n=20$)
the first order correction is roughly
120\% (200\%)
of the leading term whereas the second order contribution is 
140\% (340\%). 
This behaviour of the perturbation series 
originates from the fact that the relevant
parameter in the Coulomb system is $\as/v$ which leads to a $\as\sqrt{n}$
dependence of the moments \cite{nov:78,vz:87}. 
Thus, for higher $n$ the perturbative corrections
become increasingly more important and have to be summed up explicitly.

The resummed spectral function resulting from the imaginary part of the Green
function for the static Coulomb potential is well known \cite{ks:55,schw:73}. 
Subtracting the zero--order contribution, already included in 
$R^{(0)} = \frac{3}{2} v(3-v^2)$,
it has the form:
\bel{eq:R_C}
R_C =  \frac{9}{2}\left[\frac{x_V}{\big(1-e^{-x_V/v}\big)}-v\right]
\,, 
\ee
where $x_V\equiv \pi^2C_F a_V$ and 
$a_V$  
is the effective coupling in
the QCD potential;
$a_V$ is independent of the
renormalization scale $\mu_a$,
but it does depend on the three--momentum 
transfer between the heavy quark and antiquark,
$\vec q^{\,2}=4v^2M_b^2/(1-v^2)$.  
The expansion of $a_V$ in terms of the
$\MSb$ coupling, 
$$
a_V(\vec q^{\,2})  =  a(\mu_a)\,\left[ 1 + a(\mu_a)\,r_V^{(1)}
(\vec q^{\,2}/\mu_a^2) 
+ \ldots\right] \,,
$$
is known to $\cO(a^3)$
 \cite{fis:77,bil:80,pet:96}.          

$R_C$ resums
the leading $(a/v)^n$ and some of the sub-leading corrections \cite{vol:79}.
The corresponding terms have to be subtracted from $R^{(1)}$ and $R^{(2)}$.
Thus, we can rewrite the perturbative expansion of the spectral function
in the form
\bel{eq:4.1}
R(v)  =  S(a) \left\{ R^{(0)} + R_C +
\wt R^{(1)} a + \wt R^{(2)} a^2 \right\} \, ,
\ee
with
\begin{eqnarray} \label{eq:R_tilde} 
S(a) &\!\!\! =  &\!\!\! \Big( 1-4C_Fa+16C_F^2 a^2 \Big) ,
\no\\
\wt R^{(1)} &\!\!\! = &\!\!\! 
R^{(1)} + 4C_F R^{(0)} - \frac{9}{4}\,\pi^2 C_F \,, 
\\
\wt R^{(2)} &\!\!\! = &\!\!\! 
R^{(2)} + 4C_F R^{(1)} - \frac{3\pi^4 C_F^2}{8v} -
\frac{9}{4}\,\pi^2 C_F\, r_V^{(1)} \,.   \no
\end{eqnarray}
We have factored out in $S(a)$ the
correction to the vector current ``$-4C_F a$'',
which originates from transversal, hard gluons; 
we have added a term $16C_F^2
a^2$ in order not to generate additional corrections of $\cO(a^2)$
proportional to $R^{(0)}$. 

After performing the Coulomb resummation, the large--moment behaviour of the
remaining terms is much weaker.
For instance, although $\Mn^{(1)}/\Mn^{(0)}$ increases as
$\sqrt{n}$, now
$\wt \Mn^{(1)}/\Mn^{(0)}\sim 1/ \sqrt{n}$.
One can easily check that, if 
$a_V(\vec q^{\,2})$
is evaluated with $n_l=4$ light quark flavours,
all contributions to $R^{(2)}$
of ${\cal O}(1/v)$, ${\cal O}(\ln v^2)$ and
${\cal O}(1)$  are canceled in
\eqn{eq:R_tilde}, such that $\wt R^{(2)}$ vanishes in the limit
$v\rightarrow 0$.
Thus, to $\cO(a^2)$, all threshold singularities are properly taken into
account through the Coulomb factor $R_C$.

The contributions of
${\cal O}(v)$ which determine the constant terms in
$\wt{\cal M}_{n}^{(2)}/\Mn^{(0)}$ 
are not known analytically.    
Those terms correspond to the current
correction from transversal, hard gluons.
Thus, the splitting of the $\cO(a^2 v)$ correction between 
$\wt R^{(2)}$ and the global factor $S(a)$ is ambiguous.

\subsection{Power Corrections}

The leading non-perturbative correction is the
gluon--condensate contribution to the massive vector
correlator; it is known at the next-to-leading order
\cite{bro:94}.        

The relative growth of
${\cal M}_{n,G^2}^{(0)}/\Mn^{(0)}$ is proportional to $n^3$. Therefore, the
non-perturbative contribution grows much faster than the perturbative moments.
In addition, in the pole--mass scheme,
the next-to-leading order correction to ${\cal M}_{n,G^2}$
is of the same size or larger as
the leading term. Because the perturbative expansion for the gluon condensate
cannot be trusted, we shall restrict our analysis to the range 
$n\leq20$ where its
contribution to the $b\bar b$ moments is below 3\%.

\section{RESULTS}

\begin{figure}[bht]
\centerline{\rotate[r]{\epsfysize=7.5cm\epsffile{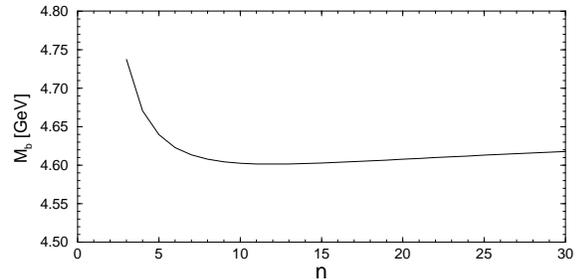}}}
\vspace{-0.5cm}
\caption{$M_b$ 
   as a function of $n$.\label{fig:1}}
\end{figure}

To suppress higher resonances as well as power corrections, we have restricted
the analysis to the range $n=8,\ldots,20$.
Solving the moment sum rules for $M_b$, we can fit $M_b$ to a constant by
varying $M_b$ and $\alpha_s(M_b)$. 
In Figure~\ref{fig:1} the resulting values for $M_b$ are displayed as a
function of $n$. This illustrates that a constant $M_b$ in the range
$8\leq n\leq 20$ really produces an excellent fit.

\begin{table}[thb]
\caption{Error estimates for  
  $M_b$ and  $\alpha_s(M_b)$. \label{tab:4}}
\vspace{0.2cm}
\begin{tabular}{lll}
\hline & $\Delta M_b\;[\mev]$ & $\Delta\alpha_s(M_b)$ \\
\hline statistical & $\pm 9.1$ & $\pm 0.0097$ \\
${\cal O}(a^3)$ Coulomb & $\pm 7.3$ & $\pm 0.0222$ \\
${\cal O}(a^2)$ & $\pm 0.9$ & $\pm 0.0082$ \\ 
scale $\mu_a$ & $\pm^{2.1}_{1.3}$ & $\pm^{0.0287}_{0.0196}$ \\ 
continuum & $\pm^{2.6}_{2.0}$ & $\pm^{0.0037}_{0.0027}$ \\
$\GG$ & $\pm 5.3$ & $\pm 0.0036$ \\
$\Gamma_{kS}$ & $\pm 3.1$ & $\pm 0.0067$ \\
\hline total & $\pm 13.5$ & $\pm 0.0269$ \\
\hline
\end{tabular}
\end{table}

 A compilation of all different contributions to the errors on $M_b$ and
$\alpha_s(M_b)$ is summarized in Table~\ref{tab:4}. The dominant
uncertainty, due to the unknown higher--order perturbative corrections,
has been estimated in three different ways:
1) the size of the known ${\cal O}(a^3)$ correction (through $a_V$
\cite{pet:96}) to $R_C$
---only the ${\cal O}(a^2)$ contribution has been included
in the central values---;
2)  the size of the ${\cal O}(a^2)$ term $\wt R^{(2)}$;
and 3) the variation under a change of the scale
at which $\alpha_s$ is evaluated, in the range
$M_b/2 \leq \mu_a \leq 2M_b$.
 Adding all three would double count the error, because the uncertainty in an
asymptotic series, such as the perturbative expansion, is bounded by the size
of the last known term. For our final results, we have chosen to include the
errors of varying the Coulomb and the ${\cal O}(a^2)$ terms.


The entry for the gluon condensate 
results from removing the non-perturbative contribution
completely.


Adding all errors in quadrature, we finally get:  
\begin{eqnarray} 
\lefteqn{M_b  =  4.604 \pm 0.014 \; \gev \,,} \label{eq:6.5} \\
\lefteqn{\alpha_s(M_b)  =  0.2197 \pm 0.0269 \label{eq:6.6} \,,}
\end{eqnarray} 
%
which implies
\begin{equation}
\label{eq:6.7}
\alpha_s(M_Z)  =  0.1184 \pm^{\;0.0070}_{\;0.0080} \,.
\end{equation} 
%


We have also investigated the same sum rules using the $\MSb$ 
quark mass.
The convergence of the perturbative series turns out to be better in the
$\MSb$ scheme. However, we have refrained from performing a resummation of the
Coulomb corrections because now the velocity
$v$ depends on the renormalization scale $\mu_m$,
used to define the running mass $m_b(\mu_m)$.
To restrict
the ${\cal O}(a^2)$ corrections to a reasonable size ($<50\% $), 
$\mu_m$ should lie in the
range $\mu_m=3.2\pm 0.5\,\gev$.

\begin{table}[thb]
\caption{Error estimates for  
  $m_b(m_b)$ and $\alpha_s(m_b)$. \label{tab:5}}
\vspace{0.2cm}
\begin{tabular}{lll}
\hline & $\Delta m_b\;[\mev]$ & $\Delta\alpha_s(m_b)$ \\
\hline statistical & $\pm 2$ & $\pm 0.0044 $ \\ 
scale $\mu_m$ & $\pm^{33}_{36}$ & $\pm^{0.0362}_{0.0058}$ \\ 
scale $\mu_a$ & $\pm^{49}_{31}$ & $\pm^{0.0251}_{0.0280}$ \\ 
continuum & $\pm 1$ & $\pm^{0.0035}_{0.0026}$ \\
$\GG$ & $\pm 2$ & $\pm 0.0023$ \\
$\Gamma_{kS}$ & $\pm 3$ & $\pm 0.0060$ \\
\hline total & $\pm^{59}_{48}$ & $\pm^{0.0449}_{0.0297}$ \\
\hline
\end{tabular}
\end{table} 

The separate contributions to the theoretical errors
have been listed in Table~\ref{tab:5}. 
The uncertainty from higher--order corrections
is now due to the variation of the scales $\mu_m=3.2\pm 0.5\,\gev$ and
$2.6\,\gev \leq \mu_a \leq 2m_b$. The scale $\mu_a$ should not be taken lower
than roughly $2.6\,\gev$ because otherwise the ${\cal O}(a^2)$ correction
$\overline{\cal M}_n^{(2)}$ becomes unacceptably large. 
Adding all errors in quadrature, we get:
\begin{eqnarray} 
\lefteqn{m_b(m_b)  =  4.13 \pm 0.06 \; \gev \,,} \label{eq:7.8} \\
\lefteqn{\alpha_s(m_b)  = 0.2325 \pm^{\;0.0449}_{\;0.0297} \,,}\label{eq:7.9} 
\end{eqnarray} 
which implies
\begin{equation}
\label{eq:7.10}
\alpha_s(M_Z)  =  0.1196 \pm^{\;0.0102}_{\;0.0080} \,.
\end{equation}

\section{DISCUSSION}

The resulting values for $\alpha_s(M_Z)$ from
the pole and $\MSb$  mass schemes turn out to be in very good agreement. 
This is
a further indication that the uncertainty from unknown higher--order 
corrections
is under control. In addition, our results $m_b(m_b)$ and
$M_b$ for the $b$--quark mass satisfy the known perturbative relation
between the pole and $\MSb$ masses \cite{tar:81,gbgs:90}, 
within the errors. 

Combining both determinations of the strong coupling constant
$\alpha_s$, we find
\begin{equation}
\label{eq:8.3}
\alpha_s(M_Z)  =  0.119 \pm 0.008 \,.
\end{equation} 
We have not averaged the errors of the two determinations because
they are not  independent.
This result is surprisingly close to the current world average \cite{bethke},
$\alpha_s(M_Z)  =  0.1186 \pm 0.0036$,
although the error is certainly larger.

Our results differ from the ones originally quoted in Ref.~\cite{vol:95},
showing that the errors were grossly underestimated. The main differences
stem from our inclusion of $\cO(a^2)$ corrections, which are large, and
from our more complete numerical analysis. (In Ref.~\cite{vol:95}, the
analysis was performed expanding in powers of $1/n$ and keeping only the
leading term; this is a quite bad numerical approximation; moreover,
the actual expansion parameter turns out to be $1/\sqrt{n}$).

The bottom quark mass values obtained by us are in good agreement to previous
determinations from QCD sum rules
\cite{rry:85,nar:89,dp:92,nar:94}   
and a very recent calculation from
lattice QCD \cite{gms:96}.
Evolving $m_b(m_b)$ to the $Z$ peak, it also agrees with the DELPHI
$m_b(M_Z)$ determination \cite{marti}.
Owing to the big sensitivity of the moment sum rules
for the $\Upsilon$ system to the quark mass, and the good control over
higher--order $\alpha_s$ corrections, our result is more precise.
This result is lower than the one obtained in potential models
\cite{Yndurain}.

The main uncertainty of our results originates in the perturbative series.
We have estimated the effect of unknown higher--order corrections
in a quite conservative way. However, it is known that the pole mass
suffers from a renormalon ambiguity  which has been argued to be of
$\cO(100 \mev)$ \cite{bb:94,bsuv:94,ns:94}.
 Throughout our analysis, the pole mass has been defined as the pole of the
perturbatively renormalized quark propagator.  Our determination \eqn{eq:6.5}
might therefore be subject to additional uncertainties which go beyond
perturbation theory but which we cannot assess in a precise way.

It has been pointed out recently \cite{Hoang} that additional Coulombic
singularities, not included in $R_C$, could show up at $\cO(a^3)$. If true,
that could generate sizeable corrections to the threshold region ($v \to 0$)
and, therefore, to the very large--n moments.
However, our results have been obtained at
relatively low values of $n$, and they are very stable in the whole range
$8\leq n\leq 20$. Thus, this kind of higher--order corrections seems to be
safely included in our quoted uncertainties.


\end{document}